\documentclass[5p]{elsarticle}
\usepackage{lineno,hyperref}
\usepackage{amsmath}
\usepackage[abbreviations]{glossaries-extra}
\usepackage[toc,page]{appendix}
\usepackage{float}
\usepackage{graphicx}
\usepackage{epstopdf}
\usepackage{subcaption}
\usepackage{pgfplots}
\pgfplotsset{compat=1.14}
\usepackage{dcolumn}
\usepackage{bm}
\usepackage{subcaption}
\hypersetup{colorlinks = true,linkcolor = black, anchorcolor =red, citecolor = blue,filecolor = red,urlcolor = red, pdfauthor=author}
\usepackage{xcolor}
\modulolinenumbers[5]
\usepackage{todonotes}
\usepackage{framed} 
\usepackage{multicol} 
\usepackage{nomencl} 
\makenomenclature
\renewcommand*\nompreamble{\begin{multicols}{2}}
\renewcommand*\nompostamble{\end{multicols}}

\usepackage{mathtools}
\usepackage{soul,color}
\soulregister\cite7
\soulregister\ref7
\soulregister\pageref7
\soulregister\gls7
\soulregister&7
\sethlcolor{yellow}

\journal{...}
\begin{document}
\begin{frontmatter}

\title{Central moments multiple relaxation time LBM for hemodynamic simulations in intracranial aneurysms: An in-vitro validation study using PIV and PC-MRI}

\author[OvGU,ETHZ]{Seyed Ali Hosseini\corref{mycorrespondingauthor}}
\cortext[mycorrespondingauthor]{Corresponding author}
\ead{seyed.hosseini@ovgu.de}
\author[OvGU,STIMULATE]{Philipp Berg}
\author[OvGU]{Feng Huang}
\author[OvGU]{Christoph Roloff}
\author[OvGU]{G{\'a}bor Janiga}
\author[OvGU]{Dominique Th\'evenin}

\address[OvGU]{Laboratory of Fluid Dynamics and Technical Flows, University of Magdeburg ``Otto von Guericke'', D-39106 Magdeburg, Germany}
\address[ETHZ]{Department of Mechanical and Process Engineering, ETH Z\"urich, 8092 Z\"urich, Switzerland}
\address[STIMULATE]{Research Campus STIMULATE, University of Magdeburg ``Otto von Guericke'', D-39106 Magdeburg, Germany}
\begin{abstract}
The lattice Boltzmann method (LBM) has recently emerged as an efficient alternative to classical Navier-Stokes solvers. This is particularly true for hemodynamics in complex geometries. However, in its most basic formulation, {i.e.} with the so-called single relaxation time (SRT) collision operator, it has been observed to have a limited stability domain in the Courant/Fourier space, strongly constraining the minimum time-step and grid size. The development of improved collision models such as the multiple relaxation time (MRT) operator in central moments space has tremendously widened the stability domain, while allowing to overcome a number of other well-documented artifacts, therefore opening the door for simulations over a wider range of grid and time-step sizes. The present work focuses on implementing and validating a specific collision operator, the central Hermite moments multiple relaxation time model with the full expansion of the equilibrium distribution function, to simulate blood flows in intracranial aneurysms. The study further proceeds with a validation of the numerical model through different test-cases and against experimental measurements obtained via stereoscopic particle image velocimetry (PIV) and phase-contrast magnetic resonance imaging (PC-MRI). For a patient-specific aneurysm both PIV and PC-MRI agree fairly well with the simulation. Finally, low-resolution simulations were shown to be able to capture blood flow information with sufficient accuracy, as demonstrated through both qualitative and quantitative analysis of the flow field {while leading to strongly reduced computation times. For instance in the case of the patient-specific configuration, increasing the grid-size by a factor of two led to a reduction of computation time by a factor of 14 with very good similarity indices still ranging from 0.83 to 0.88.}
\end{abstract}
\begin{keyword}
Lattice Boltzmann method; Particle Image Velocimetry; Intracranial aneurysm; Magnetic Resonance Imaging; Validation; {Single relaxation time; Central Hermite multiple relaxation time.}
\MSC[2010] 00-01\sep 99-00
\end{keyword}
\end{frontmatter}
\begin{table*}[!t]
   \begin{framed}
     \printnomenclature
   \end{framed}
\end{table*}
\section{Introduction}
\newabbreviation{ia}{IA}{Intracranial aneurysms}
\newabbreviation{cfd}{CFD}{computational fluid dynamics}
\gls{ia} are local malformations of the cerebral vasculature, which occur with an estimated prevalence of approximately 3\% in the western population~\cite{Vlak.2011}. It is known that aneurysms can grow and develop into a stable state or rupture, if the hemodynamic forces exceed the vascular resistance. Since a reliable and clinically applicable rupture risk assessment procedure remains challenging until now, increasing research effort is being put into developing one. Owing to their non-invasive nature resulting in risk-free (for the patient) assessment of rupture probability and the possibility to access highly resolved velocity fields (in both space and time) \gls{cfd} studies are becoming increasingly popular~\cite{Liang.2019,Saqr.2019}.\\
While valuable insights have been gained through such numerical studies~\cite{Meng.2014,Cebral.2017}, a number of challenges and gaps remain to be bridged, before predictive and reliable models based on \gls{cfd} for aneurysm dynamics can be developed~\cite{Fiorella.2011,Kallmes.2012}. To put these challenges into perspective, Berg et al.~\cite{Berg.2019} summarized the individual working steps involved in image-based blood flow simulations and provided corresponding recommendations to avoid simulation inaccuracies. As for any numerical simulation, the choice of the numerical solver, resolution and grid configuration are of the utmost importance to ensure convergence of the obtained solutions. The latter two also being consequences of the choice of the numerical method, the former is a determining factor concerning both convergence and efficiency of the solver.\\
\newabbreviation{piv}{PIV}{particle image velocimetry}
\newabbreviation{mri}{MRI}{magnetic resonance imaging}
The performances of the employed solvers can only be assessed through a systematic benchmarking/validation procedure against reliable experimental data. One of the most frequently used sources of experimental reference data is the \gls{piv} relying on laser-based high-speed camera flow measurements~\cite{Ford.2008,Raschi.2012,Yagi.2013,Bouillot.2014,Berg.2015}. \gls{piv} measurements allow for more accurate measurements and higher resolutions as compared to other data acquisition tools such as \gls{mri} \cite{van20133d} or cerebral angiography \cite{cebral2007computational}. A number of studies have presented qualitative comparisons of velocity fields for patient-specific aneurysms as obtained from \gls{cfd} and \gls{piv} measurements, and reported good qualitative agreement between the two~\cite{Paliwal.2017,Roloff.2019}.
\newabbreviation{ns}{NS}{Navier-Stokes}
\newabbreviation{nsp}{NSP}{Navier-Stokes-Poisson}
\newabbreviation{fv}{FV}{finite volume}
\newabbreviation{fe}{FE}{finite element}
\newabbreviation{sph}{SPH}{smoothed particle hydrodynamics}
\newabbreviation{lbm}{LBM}{lattice Boltzmann method}
\newabbreviation{pde}{PDE}{partial differential equation}
\newabbreviation{acm}{ACM}{artificial compressibility method}
\newabbreviation{srt}{SRT}{single relaxation time}
\newabbreviation{trt}{TRT}{two relaxation time}
\newabbreviation{mrt}{MRT}{multiple relaxation time}
\newabbreviation{chmrt}{CHMRT}{central Hermite multiple relaxation time}
\newabbreviation{edf}{EDF}{equilibrium distribution function}
\par While most of the early publications relied on discrete solvers for the \gls{nsp} equations, {e.g.} using \gls{fv} or \gls{fe} methods, emergent numerical methods such as \gls{sph}~\cite{muller2004interactive,qin2010particle} and the \gls{lbm}~\cite{zavodszky2013validation,mazzeo2008hemelb,groen2013analysing,anzai2012optimization,huang2015multi,ouared2005lattice,finck2007simulation,chen2014segmentation} are increasingly applied to such flows. Contrary to classical \gls{nsp} solvers they are not strictly incompressible, as they approximate the low Mach flow regime through appropriate isothermal flow manifolds, and as such rely on a system of purely hyperbolic equations, making the algorithm inherently local. The absence of the elliptic component of the \gls{nsp} equations allows for dramatic computation cost reduction while the parabolic nature of the \gls{pde} describing  pressure evolution makes the formulation suitable for unsteady simulations -- contrary to other weakly compressible formulations such as the \gls{acm}. Furthermore, the compressible nature of the formulation (involving a thermodynamic pressure) along with kinetic heuristic boundary closures (such as the bounce-back rule) provide for an efficient and consistent implementation of wall boundaries~{\cite{ma2018numerical,ma2019numerical,kruger2017lattice}}. The drastic decrease in computation time makes the method more viable compared to approaches based on the \gls{nsp} equations. Based on these observations, a number of dedicated \gls{lbm}-based solvers for blood flow simulations have been developed over the past decade, {e.g.}~\cite{mazzeo2008hemelb,latt2020palabos,hasert2014complex}. For example a research prototype was developed by the Siemens Healthineers AG (Erlangen, Germany), which is increasingly used to address clinical questions~\cite{Suzuki.2016,Berg.2018,Suzuki.2020}. However, it must be noted that the \gls{lbm} in its simplest form, {i.e.} with a \gls{srt} collision operator and a second-order discrete \gls{edf} has a rather limited stability domain. The \gls{srt} formulation also leads to a number of other well-documented numerical artifacts including, but not limited to, the (non-dimensional) viscosity-dependence of the solid wall position when used with bounce-back-type boundary treatments \cite{pan2006evaluation}.\\
The aim of the present work is to assess the performances of a specific class of \gls{mrt} operator based on Hermite central moments and a fully expanded \gls{edf}. This collision model is shown to alleviate some of the traditional shortcomings of the classical \gls{srt}-based solvers, while -- through its much wider stability domain -- allowing for stable low-resolution simulations. The performances of the \gls{chmrt} model are assessed, via our in-house solver ALBORZ~\cite{hosseini2019hybrid,hosseini2019theoretical}, through a variety of test-cases spanning ideal and patient-specific geometries considering steady and pulsatile flows. The numerical results are compared with high-resolution in-vitro measurements for validation. They are also compared to results from a \gls{srt} solver with second-order \gls{edf} to further showcase the added value of the \gls{chmrt} collision operator. The issue of under-resolved simulations is also considered by systematically conducting simulations at different (lower) resolutions. It is shown that at low resolution (inaccessible to the \gls{srt} collision operator), the proposed scheme is still able to capture properly the flow dynamics.
\section{Materials and methods}
\subsection{Numerical method}
\subsubsection{Lattice Boltzmann solver for the Navier-Stokes equations}
\newabbreviation{bgk}{BGK}{Batnagar-Gross-Krook}
The \gls{lbm} is a solver for the Boltzmann equation in the limit of the hydrodynamic regime~\cite{succi2001lattice}. The time-evolution of the discrete probability distribution functions is written as~\cite{kruger2017lattice}:
\begin{equation}
    f_\alpha\left(\bm{x}+\bm{c}_\alpha \delta_t,t+\delta_t\right) = f_\alpha\left(\bm{x},t\right) + \delta_t \Omega_\alpha,
\end{equation}
\nomenclature{$f_\alpha$}{Discrete probability distribution function}
\nomenclature{$\delta_t$}{Time-step size in [s]}
\nomenclature{$\bm{c}_\alpha$}{Discrete particle velocity in [m/s]}
\nomenclature{$\Omega_\alpha$}{Collision operator}
where $f_\alpha$ are the discrete distribution functions, $\bm{c}_\alpha$ the corresponding particle velocities, $\delta_t$ the time-step size and $\Omega_\alpha$ the discretized particle collision operator. The collision operator $\Omega_\alpha$ is usually approximated via a \gls{bgk} linear relaxation model defined as~\cite{bhatnagar1954model}:
\begin{equation}
    \Omega_\alpha = \frac{1}{\tau}\left(f_\alpha^{(eq)}-f_\alpha\right),
\end{equation}
\nomenclature{$\tau$}{Relaxation time in [s]}
\nomenclature{$f_\alpha^{(eq)}$}{Discrete equilibrium distribution function}
where $f_\alpha^{(eq)}$ is the \gls{edf}. The \gls{edf} used in \gls{lbm} solvers is a truncated approximation (using a Taylor-McLaurin expansion in the limit of vanishing Mach numbers~\cite{he1997theory} or a Hermite expansion~\cite{shan2006kinetic}) to the Maxwell-Boltzmann distribution. It can be written as~\cite{hosseini2019stability}:
\begin{equation}\label{eq:EDF}
    f_\alpha^{(eq)} = w_\alpha \sum_{n=0}^{N} \frac{1}{n!c_s^{2n}} \bm{a}^{(eq)}_n:\bm{\mathcal{H}}_{\alpha,n},
\end{equation}
\nomenclature{$\bm{\mathcal{H}}_{\alpha,n}$}{Hermite polynomial of order $n$}
\nomenclature{$\bm{a}^{(eq)}_n$}{Hermite equilibrium coefficient of order $n$}
\nomenclature{$w_\alpha$}{Weight associated to discrete velocity $\bm{c}_\alpha$}
\nomenclature{$c_s$}{lattice sound speed in [m/s]}
where $w_\alpha$ are weights associated to each discrete population, $c_s$ is the non-dimensional sound speed at the reference temperature tied to the time-step and grid sizes, $\bm{\mathcal{H}}_{\alpha,n}$ the Hermite polynomial of order $n$ and $\bm{a}^{(eq)}_n$ the Hermite coefficient of the corresponding order. Conserved field variables appearing in the \gls{edf}, {i.e.} density and momentum, are computed from distribution functions as:
\begin{equation}
    \rho = \sum_\alpha f_\alpha,
\end{equation}
\begin{equation}
    \rho \bm{u} = \sum_\alpha \bm{c}_\alpha f_\alpha,
\end{equation}
where $\rho$ and $\bm{u}$ are the fluid density and velocity.
\nomenclature{$\rho$}{Density in [kg/$\hbox{m}^3$]}
\nomenclature{$\bm{u}$}{Velocity in [m/s]}
\subsubsection{The \gls{srt} formulation: shortcomings}
As briefly mentioned in the introduction, the form of the \gls{lbm} most commonly used in medical applications, {i.e.} with the \gls{srt} collision operator and a second-order \gls{edf}, is subject to a number of well-documented shortcomings. It can be readily shown through a multi-scale perturbation analysis, that at the macroscopic scale, the viscous stress tensor recovered by the second-order \gls{edf} admits deviations from the target Newtonian stress scaling as $\mathcal{O}(\hbox{Ma}^3)$~\cite{kruger2017lattice}. In the \gls{lbm} literature this deviation is usually referred to as the Galilean invariance problem (rightly so as it ties the effective kinematic viscosity to local velocity) of the stress tensor. It can readily be shown that the Galilean invariance of the viscous stress tensor can be restored by extending the Hermite expansion of the \gls{edf} to higher orders (at least order three). For the bulk viscosity however, given the bias between the first and third-order moments introduced by the limited number of discrete velocities, one must introduce an additional correction term. The bulk viscosity not being of much interest nor relevance in the near-incompressible regime, it will not be discussed further. Detailed numerical and theoretical proofs can be found in~\cite{prasianakis2008lattice,karlin2010factorization,hosseini2020compressibility,HosseiniPhD2020,saadat2019lattice}.\\
One of the major advantages of the \gls{lbm} is the way boundary conditions can be implemented. The bounce-back rule is one of the most popular approaches to enforce wall, velocity and pressure boundary conditions. However, it has been shown that when used with the \gls{srt} operator it is subject to numerical artifacts such as the (non-dimensional) viscosity-dependence of the solid wall position~\cite{pan2006evaluation}. This shortcoming can be readily proven through asymptotic analysis of the corresponding system of discrete equations or simple permeability studies of flow in porous media~\cite{hosseini2019theoretical,kruger2017lattice}. The seminal work of I.~Ginzburg and subsequent development of a collision operator with a wall position independent from the relaxation coefficient led to the \gls{trt} operator~\cite{ginzburg2008two}. The so-called generalized collision operator first proposed in \cite{higuera1989lattice,d1992generalized} was later parametrized to fix its shortcomings~\cite{pan2006evaluation}. It has been shown that in order to control the wall position independently from the viscosity, at least two different relaxation coefficients are needed (one for odd, one for even moments).\\
Last but not least, it is common knowledge that the classical \gls{srt} model is very sensitive to the Fourier number defined as $\hbox{Fo}=\frac{\nu \delta_t}{\delta_x^2}$. It is practically unusable for $\hbox{Fo}<0.005$ 
as shown by \cite{hosseini2019stability,HosseiniPhD2020,lallemand2000theory}. Extension of the operation range of the \gls{lbm} has been the topic of a wide number of articles over the past decades. Continuous effort has been dedicated to developing more advanced collision operators resulting in a plethora of models such as the \gls{mrt} models based on either raw~\cite{d2002multiple} or central moments~\cite{geier2006cascaded}, regularized models~\cite{latt2006lattice}, entropic operators \cite{ansumali2002single,karlin2014gibbs,bosch2015entropic}, each of which have been shown to lift this restriction to different extents.\\
The \gls{mrt} collision operator in central Hermite moments space with a full Hermite expansion of the \gls{edf} has been shown to have better numerical properties than other choices of moments space. It provides the additional degrees of freedom necessary to fix the wall position in the bounce-back algorithm. The enhanced spectral properties of the collision operator do not only remove some of the issues inherent to second-order \gls{edf}-based \gls{srt} operators, but can effectively allow for cheaper simulations of a given configuration as they can allow for a reduction of the number of grid-points and discrete time-steps.
\subsubsection{\gls{chmrt} formulation}
The classical \gls{bgk} collision operator with one relaxation coefficient having stability issues for vanishing non-dimensional viscosities, in the context of the present study we use the \gls{chmrt} model defined as:
\begin{equation}
    \Omega_\alpha = \bm{\mathcal{T}}^{-1}\bm{S} \bm{\mathcal{T}} \left(\bm{f}^{(eq)} - \bm{f}\right),
\end{equation}
\nomenclature{$\delta_x$}{Grid size in [m]}
\nomenclature{$\bm{\mathcal{T}}$}{Central Hermite moments transform matrix}
\nomenclature{$\widetilde{\bm{a}}^{(eq)}_n$}{Central Hermite equilibrium coefficient of order $n$}
where $\bm{\mathcal{T}}$ is the moments transform matrix, and $\bm{S}$ the moments relaxation rate diagonal matrix. In the chosen moments space, assuming a D3Q27 stencil, the following equilibrium moments are recovered:
\begin{equation}
    \bm{\mathcal{T}} \bm{f}^{(eq)} = \widetilde{\bm{a}}^{(eq)},
\end{equation}
where $\widetilde{\bm{a}}^{(eq)}$ are the equilibrium Hermite coefficients in central moments space:
\begin{subequations}
\begin{alignat}{4}
		\widetilde{\bm{a}}^{(eq)}_0 &= \rho, \\
		\widetilde{\bm{a}}^{(eq)}_n &= 0, \forall n\neq 0.
\end{alignat} \label{eq:Hermite_coeff_eq}
\end{subequations}
It must be noted that these equilibrium moments are only recovered using the full Hermite expansion supported by the stencil, and not the classical second-order \gls{edf} usually employed in \gls{lbm} solvers. Given the added benefits of a fully expanded \gls{edf}~\cite{hosseini2019stability,HosseiniPhD2020}, the present work will use a sixth-order Hermite-expansion for all simulations.\\
Apart from the relaxation rates of second-order moments, {i.e.} $a_{xx}$, $a_{yy}$, $a_{zz}$, $a_{xy}$, $a_{xz}$ and $a_{yz}$, tied to the fluid viscosity, the remainder of the relaxation rates are freely-tunable parameters. Appropriate choices of these free parameters have been shown to widen the linear stability domain of the solver and reduce dispersion errors~\cite{hosseini2020compressibility,HosseiniPhD2020}.
\subsubsection{Curved boundary conditions}
{Given the complex nature of geometries considered in aneuryms, appropriate treatment of wall boundaries and their curvature is an important point. In the context of the present study the curved-boundary bounce-back formulation proposed in~\cite{bouzidi2001momentum} is used. At a given boundary node $\bm{x}_f$, the missing incoming populations are computed as:}
\begin{subequations}
	\begin{align}
		f_\alpha(\bm{x}_f, t+\delta_t) &= 2qf_{\bar{\alpha}}(\bm{x}_f+\bm{c}_{\bar{\alpha}}, t+\delta_t)\nonumber \\  &+\left(1-2q\right)f_{\bar{\alpha}}(\bm{x}_f, t+\delta_t), \forall q<\frac{1}{2},\\
		f_\alpha(\bm{x}_f, t+\delta_t) &= \frac{1}{2q}f_{\bar{\alpha}}(\bm{x}_f+\bm{c}_{\bar{\alpha}}, t+\delta_t)\nonumber \\  &+\frac{2q-1}{2q}f_{\bar{\alpha}}(\bm{x}_f, t+\delta_t), \forall q\geq\frac{1}{2},
		\end{align}
	\label{Eq:CE_moments_eq}
\end{subequations}
{where $\bar{\alpha}$ designates the direction opposite $\alpha$ and $q$:}
\begin{equation}
    q = \frac{\lvert\lvert \bm{x}_f - \bm{x}_s\lvert\lvert}{\lvert\lvert\bm{c}_\alpha \lvert\lvert},
\end{equation}
{with $\bm{x}_s$ denoting the wall position in direction $\alpha$.}
\subsection{Cases description}
To illustrate the points brought forward in the previous section and showcase the performances of the \gls{chmrt} collision operator for the applications of interest, three different test-cases are considered: (a) ideal aneurysm under steady flow, (b) ideal aneurysm subject to pulsatile flow, and finally (c) patient-specific geometry. While the latter two are intended as detailed validations of the solver against experimental data, the first test-case is presented to briefly illustrate the difference between the classical \gls{srt} operator and the proposed model.
\subsubsection{Idealized aneurysm model under steady flow}
The first configuration is a rather simple one consisting of an ideal spherical aneurysm of radius 20~mm and vessels of diameter 6~mm positioned with an angle of 90$^{\circ}$. The distance between the inlet and the opposite point in the aneurysm sac is 60~mm while for the outlet it is 50~mm. The domain is subject to a constant inflow at the inlet and an open boundary at the outflow. The case is modeled using both the \gls{chmrt} and second-order \gls{srt} collision operators at different resolutions. The obtained results are further validated against the commercial finite-volume code STAR-CCM+ 14.04 (Siemens Product Lifecycle Management Software Inc., Plato, TX, USA).
\subsubsection{Idealized aneurysm model with pulsatile flow}
\begin{figure*}[!ht]
	\centering
	\hspace{-0.75\textwidth}
	\begin{subfigure}{0.25\textwidth}
		\includegraphics{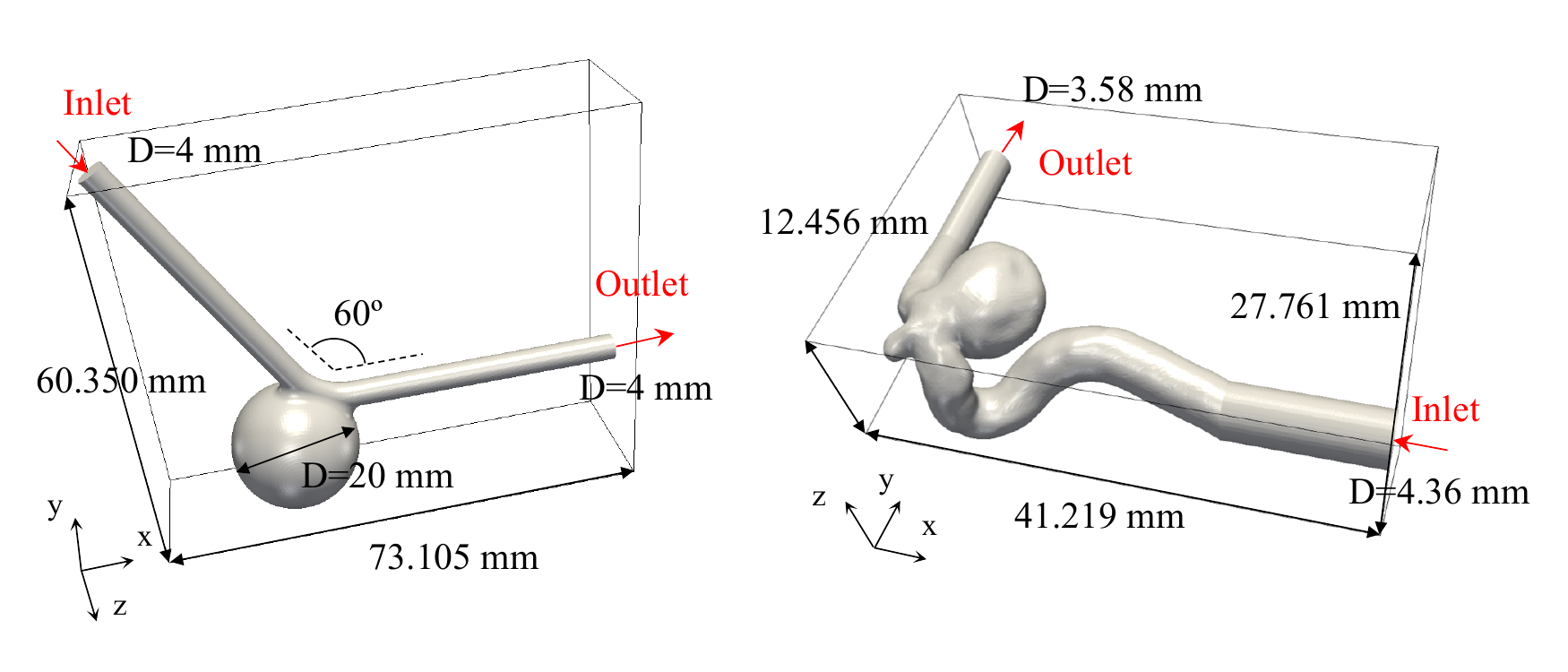}
	\end{subfigure}
	\caption{(left) Idealized aneurysm geometry as used both for the phantom manufacturing and subsequent \gls{piv} measurement and in LB simulation~{\cite{gaidzik2019transient}}, and (right) Segmentation of the patient-specific intracranial aneurysm model used for the \gls{lbm} simulations and the in-vitro validation~{\cite{roloff2019comparison}}.}
	\label{fig:geometries}
\end{figure*}

\begin{figure}[!ht]
	\centering
	\hspace{-0.1\textwidth}
	\begin{subfigure}{0.3\textwidth}
		\includegraphics{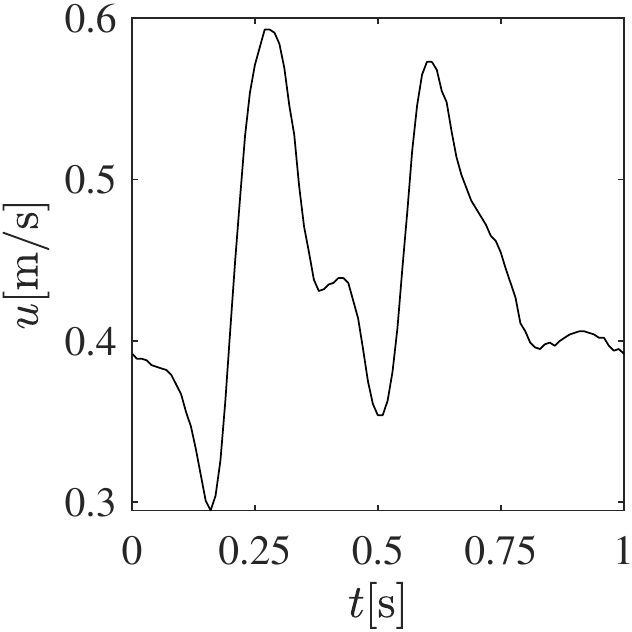}
	\end{subfigure}
	\caption{Average velocity profile at the inlet over one cardiac cycle for the idealized pulsatile aneurysm.}
	\label{fig:case03_inlet_velocity_cycle}
\end{figure}

For the purpose of validation, an idealized spherical sidewall aneurysm with a diameter of 20~mm was virtually created. The parent vessel featured a diameter of 4~mm and was bended under an angle of 120$^{\circ}$ (see Fig.~\ref{fig:geometries}). The 3-D vessel geometry served as basis for the numerical simulation as well as for the phantom manufacturing prior to the in-vitro \gls{piv} measurements.
\newabbreviation{bal}{BAL}{blood analogue liquid}
\newabbreviation{pcmri}{PC-MRI}{phase contrast magnetic resonance imaging}
Based on the 3-D CAD model, a transparent phantom model was manufactured using a lost-core technique that resulted in a silicone block incorporating the hollow vessel structure. The refractive index of the silicone was measured to be $n_{\hbox{\footnotesize silicone}}=1.4113$ at ${22}^{\circ}$C (Abbemat 200, Anton Paar, Ostfildern, Germany). As  \gls{bal}, a mixture of distilled water, glycerin, sodium iodide, and sodium thiosulphate was used to match the refractive index of the silicone block ($n_{\hbox{{\footnotesize BAL}}}=1.4109$) as well as relevant fluid dynamical properties of blood plasma, {i.e.} density $\rho_{\hbox{\footnotesize BAL}}= 1221~\hbox{kg/m}^3$. The kinematic viscosity of the fluid used for this configuration was $\nu_{\hbox{\footnotesize BAL}}= 3.2\times 10^{-6}~\hbox{m}^2/\hbox{s}$. This corresponds to a dynamic viscosity of $\mu_{\hbox{\footnotesize blood}}=4.03\times 10^{-3}~\hbox{Pa.s}$ (assuming a density of $\rho_{\hbox{\footnotesize blood}}=1222~\hbox{kg/m}^3$).\\
During the \gls{piv} measurements, the phantom was placed inside a transparent acrylic box with two inclined walls and filled with index-matching fluid. A laser light sheet was directed to illuminate the sagittal plane of the aneurysm. As seeding for the \gls{piv} measurements, small resin microspheres doped with Rhodamine B (diameter d=$10.46 \pm 0.18~\mu$m, density  $\rho=1510~\hbox{kg/m}^3$) were used. The two stereoscopic \gls{piv} high-speed cameras (sCMOS, $2560\times2160$ pixel) observed the flow through the inclined windows from the side of the acrylic box to minimize optical aberrations such as astigmatism.\\
A micro-gear pump (HNP Mikrosysteme, Schwerin, Germany) delivered the periodic flow, which was monitored by an ultrasonic flowmeter (Sonotec, Halle, Germany). The average velocity at the inflow cross section is displayed in Fig.~\ref{fig:case03_inlet_velocity_cycle} over one entire cycle (maximum Reynolds number $\hbox{Re}_{\hbox{\footnotesize max}}=1025$ and Womersley number $\hbox{Wo}=2.54$). The original flow curve is based on 4-D flow measurements in a 7T \gls{pcmri} -- Siemens Magnetom. Further details regarding the flow acquisition can be found in~\cite{berg2014validation}. To ensure a fully developed laminar flow profile the inlet connector to the phantom consisted of a 500~mm straight tubing. The \gls{piv} double frame recordings were conducted at a frequency of 500~Hz (triggered by the pump control), where the interframe time was set to $200~\mu\hbox{s}$. In total, 36 periodic cycles were recorded, resulting in 36,000 double frame pairs.
Velocity processing was conducted via DaVis 8.4.0 (LaVision, Göttingen, Germany) using a multi-pass stereo cross-correlation with a final interrogation window size of $32\times32$ px and 50~\% overlap, resulting in one velocity vector every $141~\mu\hbox{m}$. The velocity fields were then phase-averaged. Additional details on the \gls{piv} data acquisition and processing procedure can be found in~\cite{gaidzik2019transient}.
\par
\begin{figure}[!ht]
	\centering
	\hspace{-0.1\textwidth}
	\begin{subfigure}{0.3\textwidth}
		\includegraphics{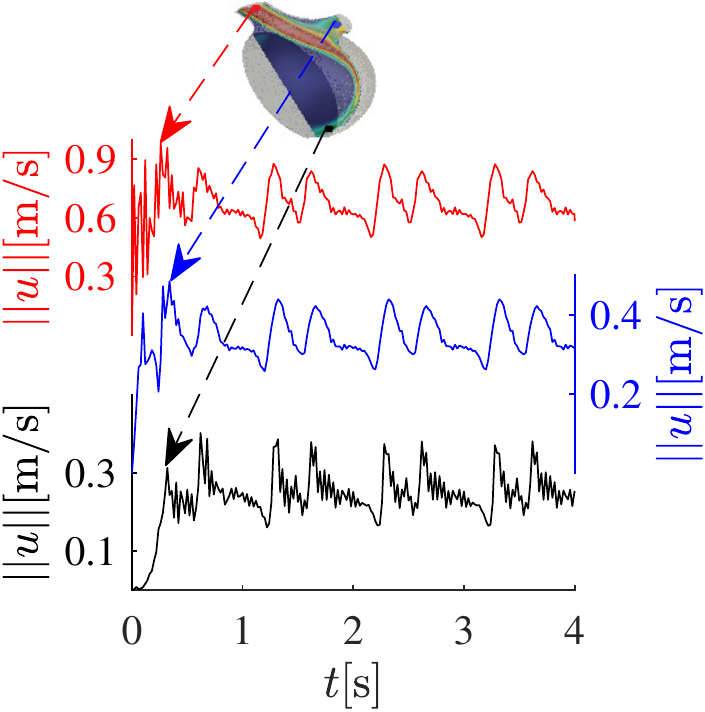}
	\end{subfigure}
	\caption{Velocity magnitude changes over four cycles at three different monitoring points around the aneurysm sac for the idealized pulsatile aneurysm.}
	\label{figure:case03_monitoring_state}
\end{figure}

 A pulsatile hemodynamic simulation was carried out using identical conditions compared to the experiment. Specifically, the same geometry and boundary conditions were applied, with a higher temporal and spatial resolution. The \gls{lbm} simulation was performed using time-step and grid sizes of $\delta_t = 1.5 \times 10^{-5}~\hbox{s}$ and $\delta_x = 1.25 \times 10^{-4}~\hbox{m}$, respectively. Two additional lower resolution simulations were also conducted to provide qualitative/quantitative characterizations of convergence and under-resolution effects. Furthermore, the simulations were ran for multiple cycles and data sampling was carried out at the fourth cycle. The state of the flow, and the absence of any artifacts tied to initial conditions were assessed by monitoring the velocity changes at three points around the aneurysm sac. The obtained results (from the highest resolution simulation) are shown in Fig.~\ref{figure:case03_monitoring_state}.
\subsubsection{Patient-specific aneurysm model with steady inflow}
In order to increase the anatomical complexity and enable a qualitative and quantitative comparison in a more realistic scenario, a patient-specific aneurysm was further considered (see Fig.~\ref{fig:geometries}). Here, an internal carotid artery aneurysm with a spherical shape was chosen, since the location represents one of the most frequent sites of occurrence. Further, a small aneurysm is located opposite the larger one leading to the formation of interesting flow structures.\\
To account for the most comprehensive situation from a hemodynamic perspective, peak systolic flow conditions were chosen for the validation. Specifically, a flow rate of $430~\hbox{mL/min}$ was defined resulting in a Reynolds number of $\hbox{Re}_{\hbox{\footnotesize peak}} = 563$ and a corresponding mean velocity in the proximal vessel of $0.512~\hbox{m/s}$, respectively.\\
Equivalent to the experiments described in the previous section, stereoscopic \gls{piv} measurements were carried out to obtain in-vitro flow data. Specifically, the micro-gear pump (HNP Mikrosysteme, Schwerin, Germany) provided the desired flow rate of $Q=430.4 \pm 1.5~\hbox{mL/min}$ and 500 recordings were taken at 5~Hz. The velocity processing was conducted via DaVis 8.4.0 (LaVision, Göttingen, Germany), which included a multi-pass cross-correlation with a final interrogation window size of $32\times32$~px (corresponding to $208\times208 ~\mu\hbox{m}$) and 50 \% overlap. This resulted in a velocity vector every 104 $~\mu\hbox{m}$. The final results were obtained by averaging all processed recordings.\\
In addition to the \gls{piv} acquisitions, phase-contrast \gls{pcmri} measurements were carried out in a 7 Tesla whole-body system (Siemens Healthineers, Forchheim, Germany) with a voxel size of $570\times570\times570~\mu\hbox{m}^3$. The measurements were repeated six times and afterwards the data were averaged. Before evaluation, eddy currents and other background phase effects were corrected using a second dataset with identical scan parameters. For further details regarding the scan parameters and the data processing the interested reader is referred to~\cite{Roloff.2019,Roloff.2017}.
\par For the corresponding \gls{lbm} simulation, similar to the idealized geometry, blood is assumed to be a Newtonian fluid with density $\rho  = 1221~\hbox{kg/m}^3$ and kinematic viscosity $\nu = 3.2\times 10^{-6}~\hbox{m}^2/\hbox{s}$, a valid choice since shear-thinning effect in the considered neurovasculature are negligible. The time-step and grid sizes were set to $\delta_t = 5 \times 10^{-6}~\hbox{s}$ and $\delta_x = 1 \times 10^{-4}~\hbox{m}$, respectively. This choice of parameters led to a non-dimensional relaxation coefficient of $\frac{\tau}{\delta_t}=0.506$ and a maximum non-dimensional velocity of $\frac{u_{\hbox{\scriptsize max}}\delta_t}{\delta_x}\approx 0.064$. To further highlight the added value of the collision operator for under-resolved flows, an additional low resolution simulation with $\delta_t = 1 \times 10^{-5}~\hbox{s}$ and $\delta_x = 2 \times 10^{-4}~\hbox{m}$, leading to $\frac{\tau}{\delta_t}=0.503$, was performed. It is worth noting that using the \gls{srt} collision operator at this resolution led to unstable simulations, as confirmed by linear stability maps found in the literature~\cite{hosseini2019stability}.
\section{Results}
In the following, qualitative and quantitative comparisons between the hemodynamic simulations based on the LB approach and the in-vitro phantom measurement are presented.
\subsection{\gls{srt} vs \gls{chmrt}: Idealized aneurysm with steady flow-rate}
\begin{figure*}[!ht]
		\begin{subfigure}{0.3\linewidth}
			\centering
			\includegraphics[width=\linewidth]{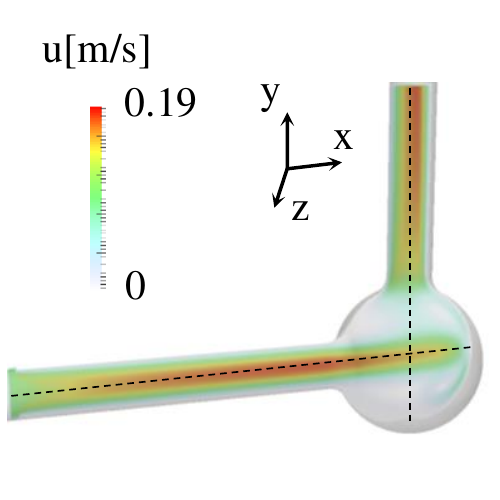}
		\end{subfigure}
		\begin{subfigure}{0.35\linewidth}
			\centering
			\includegraphics[width=0.8\linewidth]{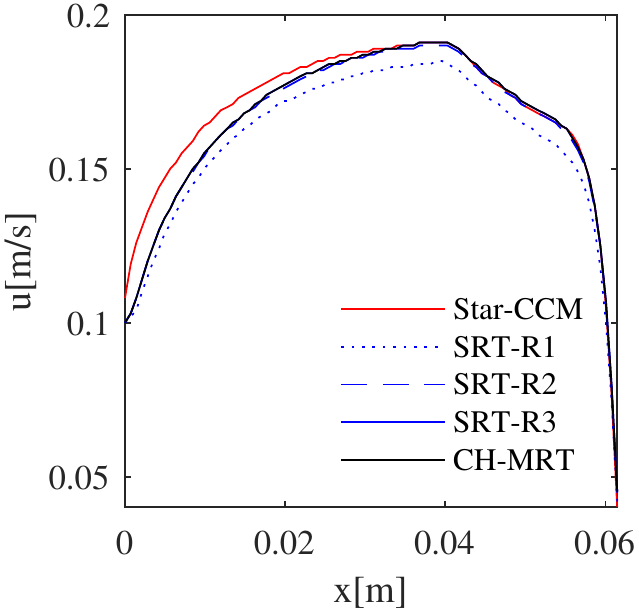}
		\end{subfigure}
		\hspace{-0.6cm}
		\begin{subfigure}{0.35\linewidth}
			\centering
			\includegraphics[width=1.1\linewidth]{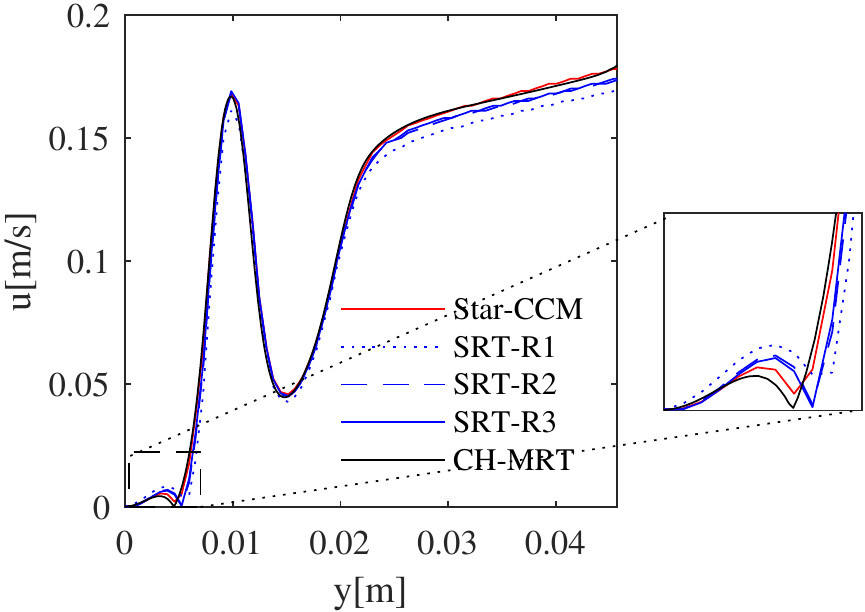}
		\end{subfigure}	
		\caption{(from left to right) Geometrical configuration of the steady flow in the idealized aneurysm, velocity distribution in the $x-$direction, and velocity distribution in the $y-$direction (along the dashed lines shown in the left subfigure).}
		\label{fig:ideal_steady_aneurysm}
	\end{figure*}

As a first step, before going into detailed validation of the solver against experimental data, we present a brief comparative study of the solver (against a classical \gls{srt} solver) through a simple test-case. The steady-state velocity profiles along the incoming and outgoing vessels, as obtained from the different simulations are shown in Fig.~\ref{fig:ideal_steady_aneurysm}. The simulations using the classical \gls{srt} model were conducted using three different resolutions, {i.e.} $\delta_x=3.33\times10^{-4}$, $2.22\times10^{-4}$ and $1.11\times10^{-4}$~m (R1, R2, and R3 in Fig.~\ref{fig:ideal_steady_aneurysm}). For the \gls{chmrt} only one simulation at $\delta_x=3.33\times10^{-4}$ was conducted.
It can clearly be observed that at the lowest resolution, the \gls{chmrt} solver matches results from the \gls{srt} simulation with the highest resolution, and even surpasses it in some regions. The low-resolution \gls{srt} simulation exhibits clear underestimation of the maximum velocity at the center of the inflow vessel. The \gls{chmrt} simulation on the other hand, even through relying on a grid-size three times that of the highest resolution \gls{srt}, is the closest to the reference STAR-CCM+ simulation, showing the superior performances of the former. It is worth noting that the differences observed near the inlet are to be expected, due the different ways boundary conditions are applied. The length of the inlet pipe, as observed in the velocity distribution plots, guarantees that at the inlet of the sac the velocity profiles are established for all simulations.
\subsection{Idealized aneurysm model with pulsatile inlflow}
\begin{figure*}[!ht]
	\centering
	\hspace{-0.75\textwidth}
	\begin{subfigure}{0.3\textwidth}
		\includegraphics{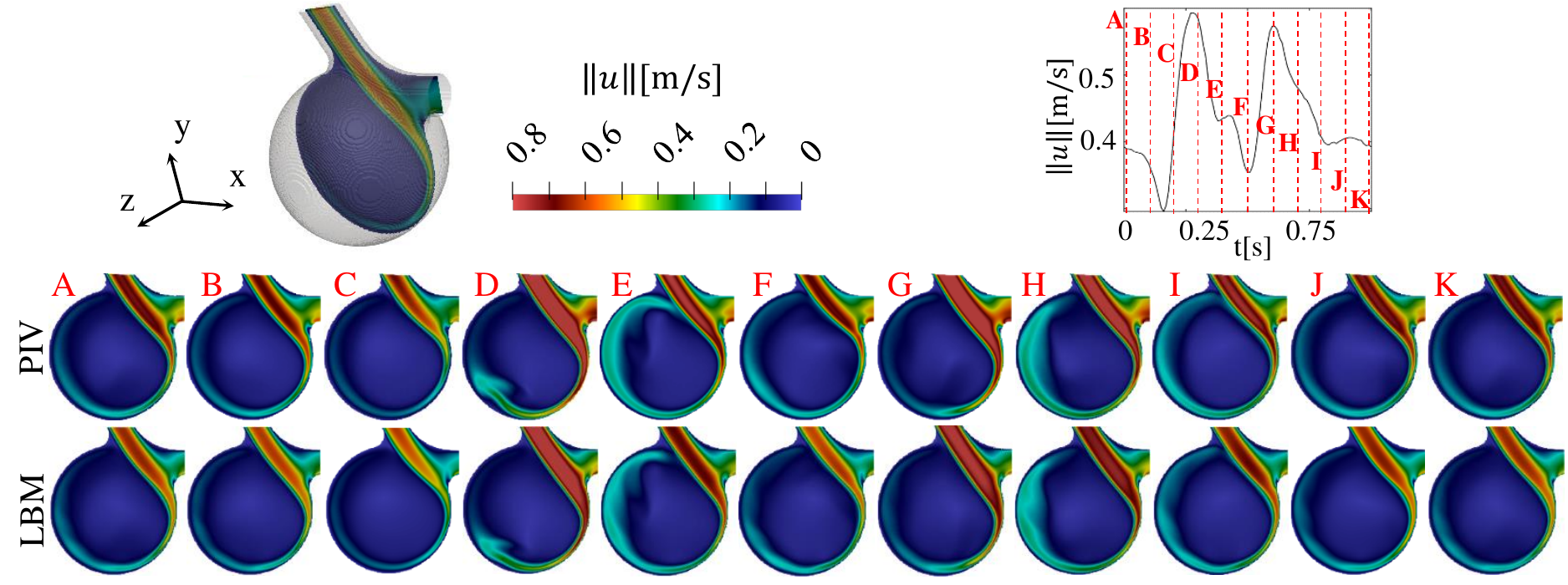}
	\end{subfigure}
	\caption{Qualitative comparison of the transient flow field between \gls{piv} experiments (top row) and LB simulation (bottom row) at eleven different time points covering an entire cardiac cycle (from 0~s to 1~s with $\Delta t$=0.1~s). The considered plane is shown on the upper left figure. The positions of the snapshots in time (relative to the cardiac cycle) are shown with red dashed lines on the upper right plot.}
	\label{fig:case03_velocity_fields}
\end{figure*}

\begin{figure*}[!ht]
	\centering
	\hspace{-0.98\textwidth}
	\begin{subfigure}{0.01\textwidth}
		\includegraphics{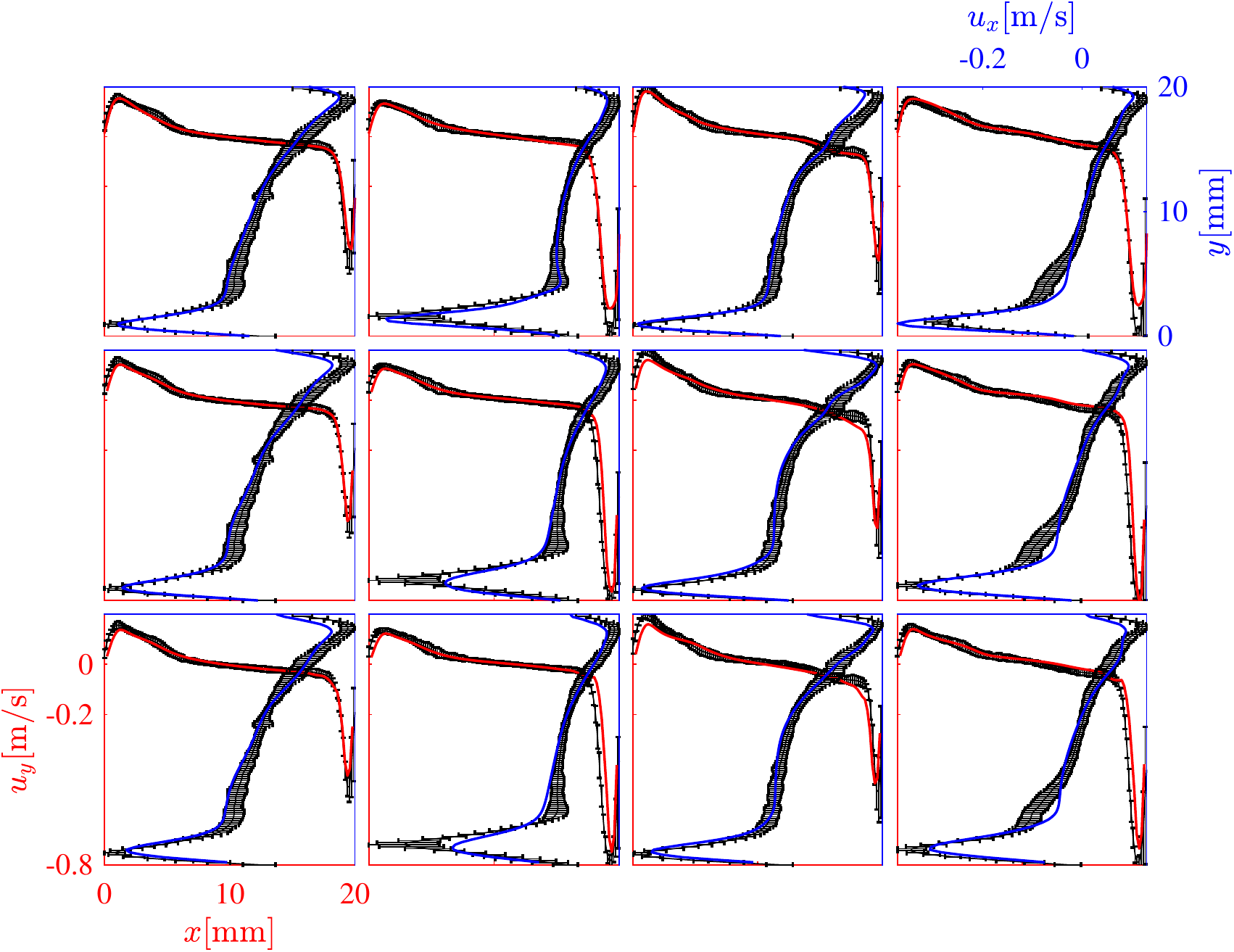}
	\end{subfigure}
	\caption{Velocity profiles along the (in red) $x$- and (in blue) $y$-directions at the center of the aneurysm sac on the central $z$-plane at four different times: (from left to right) 0.16, 0.27, 0.51 and 0.61 s from the start of the fourth cycle, and (top to bottom) decreasing resolutions in the lattice Boltzmann simulations. Velocity profiles obtained from the \gls{lbm} simulation are shown with red and blue plain lines while black plain lines with error bars (corresponding to the 99.7\% confidence interval) represent data obtained from \gls{piv} measurements.}
	\label{fig:case03_velocity_profiles}
\end{figure*}

The angle of the model for the second case was chosen in a way that the flow can enter into the sac through the distal part of the aneurysm ostium. After impinging on the aneurysm wall, the flow aligns along the perfect spherical shape. This forms a large vortex within the aneurysm with a stagnation zone approximately at the center of the sac.\\
The comparison between the experimental (\gls{piv}) and the numerical (\gls{lbm}) flow acquisition demonstrates an excellent agreement within the considered plane (perpendicular to the $z$-axis passing through the center of the aneurysm sac). Results for 11 different times spanning a cycle are shown in Fig.~\ref{fig:case03_velocity_fields}. Almost identical flow structures are captured and slight deviations are only visible in the outflow region of the aneurysm close to the peak-systolic time point (0.4~s). Additionally, minor differences can be observed most notably in the form of, less pronounced, unsteady structures not observed in the experimental data. The absence of these structure can partially be explained by the data acquisition modes in experiments and simulations. While the time-dependent flow field from the \gls{piv} are phase-averaged over a number of cycles, those from the simulations represent instantaneous shots from the fourth cycle after initialization.
In addition to this qualitative observation, a quantitative comparison is presented in Fig.~\ref{fig:case03_velocity_profiles}. Here, the velocity components ($u_x$ and $u_y$) are compared along two orthogonal lines on the plane perpendicular to the $z$-axis for three different resolutions, confirming the initial findings: Both experiment and simulation lead to almost identical profiles throughout the cardiac cycle and the numerical approach nearly always occurs within the presented error bars of the measurement. It is worth noting that two additional simulations were performed by setting $\delta_x$ to $2\times10^{-4}$~m and $3\times10^{-4}$~m; the corresponding time-steps were set to respectively $1.5\times10^{-5}$~s and $2.2\times10^{-5}$~s. At the lowest resolution, as compared to the highest resolution simulation, the computational cost is reduced by a factor of 25, while velocity profiles are still in very good agreement (and mostly within the error bars) with the experimental data.
Overall, based on both qualitative and quantitative comparisons with \gls{piv} data, it is shown that the solver is able to correctly capture the flow structure and match experimental observations. The good agreement is still well maintained even for relatively under-resolved simulations.
\subsection{Patient-specific aneurysm model}
\begin{figure*}[!ht]
	\centering
	\hspace{-0.5\textwidth}
	\begin{subfigure}{0.3\textwidth}
		\includegraphics{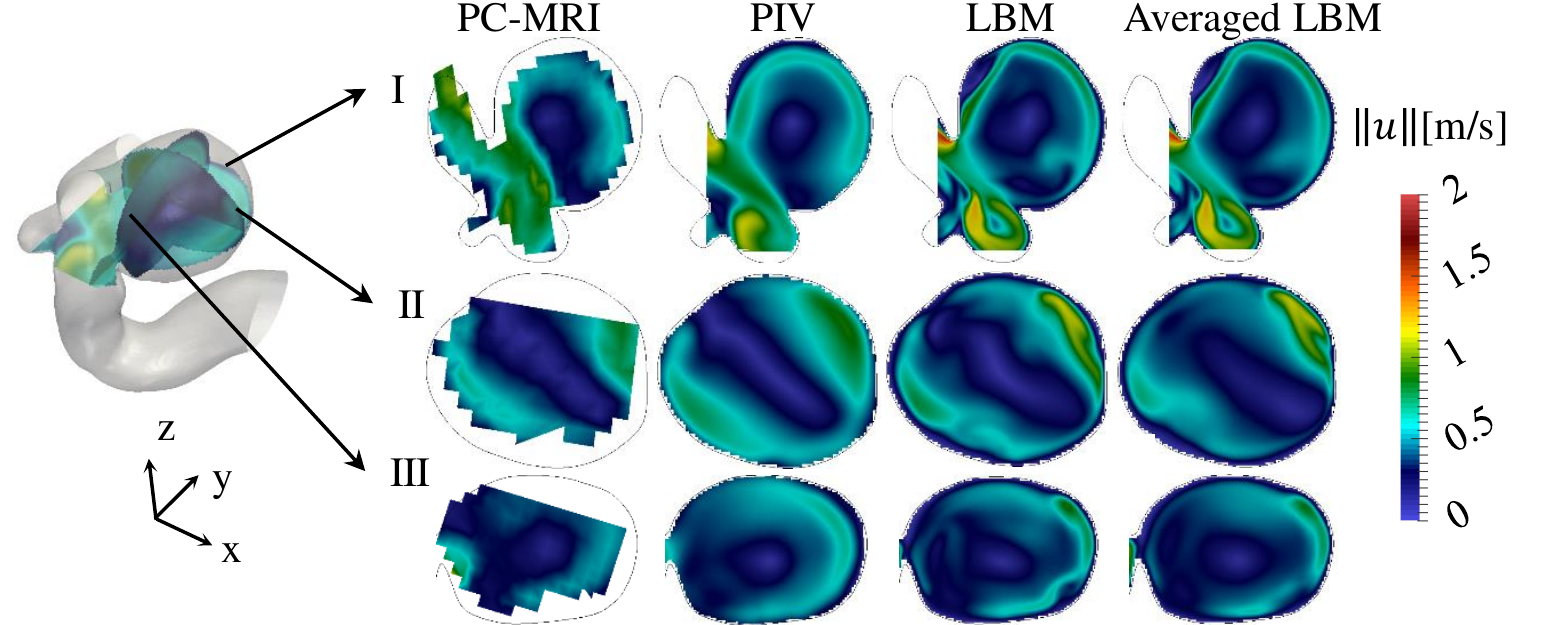}
	\end{subfigure}
	\caption{From left to right: Qualitative comparison of the velocity fields acquired using \gls{pcmri} (1st column), stereoscopic \gls{piv} (2nd column), instantaneous (3rd column) and time-averaged \gls{lbm} (last column) simulation in the aneurysm sac on three different planes, {i.e.} from top to bottom: planes I, II and III.}
	\label{fig:case02_velocity_fields}
\end{figure*}

\begin{figure*}[!ht]
	\centering
	\hspace{-0.65\textwidth}
	\begin{subfigure}{0.3\textwidth}
		\includegraphics{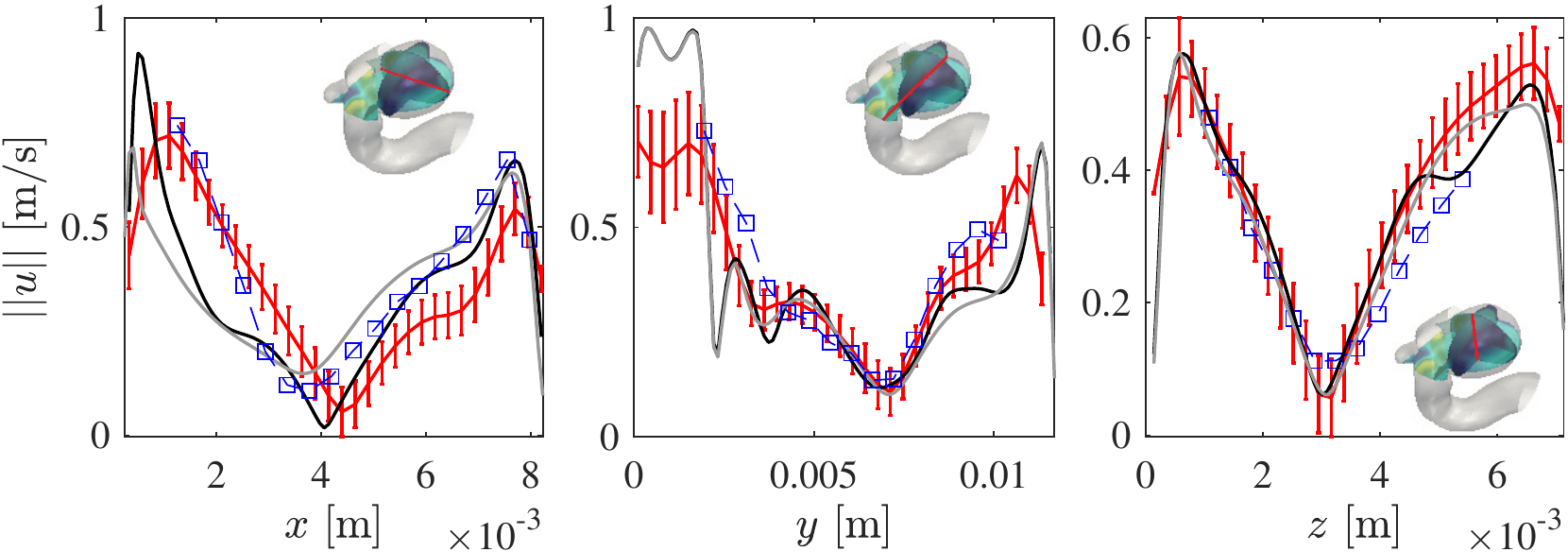}
	\end{subfigure}
	\caption{Comparison of velocity magnitude profiles along three lines in the aneurysm sac as obtained from \gls{piv} (red plain lines with error bars corresponding to one standard deviation), \gls{pcmri} (blue dashed lines with square symbols), instantaneous LB velocity field (black plain lines) and time-averaged (over 0.25~s) LB velocity field (grey plain lines).}
	\label{fig:case02_velocity_profile}
\end{figure*}

The last part of the validation was carried out in a patient-specific aneurysm model enabling a clearly more realistic flow scenario. The obtained velocity fields (both numerical and experimental) on three perpendicular planes cutting through the main aneurysm sac are shown in Fig.~\ref{fig:case02_velocity_fields}. It must be noted that both \gls{piv} and \gls{pcmri} measurements were conducted here (as opposed to the previous configuration where only \gls{piv} measurements were made).\\
Similar to the idealized test-case, a vortex forms within the sac aligning along the luminal surface and forming a stagnation zone at the center. Due to the helical vessel anatomy proximal to the aneurysm as well as the different ostium size, this vortex appears less stable.\\
Regarding the qualitative comparison of the three independent methods of flow acquisition one can notice that the overall flow structure is captured by all approaches. The entering flow jet of the simulation appears to be slightly narrower and contains accordingly higher velocity values compared to the experimental techniques. A look at the convergence indicator in the simulations and the flow field evolution over time showed that this configuration results in an unsteady flow although the inlet velocity is not pulsatile. This numerical observation was also confirmed by studying \gls{piv} instantaneous velocity fields. As such the data averaging procedure in time, {i.e.} sampling frequency and time interval, can have non-negligible effect on out-coming velocity fields from both experiments and numerical simulations. The third column in Fig.~\ref{fig:case02_velocity_fields}, representing the instantaneous fields from the \gls{lbm} simulation are included to point out the unsteady nature of the flow. It must be noted that due to the computational costs and large sizes of associated files, an interval of 0.25~s and sampling frequency of 400~Hz were used to get the average velocity fields for the simulation while as noted earlier, for the \gls{piv} the data acquisition was operated at a frequency of 5~Hz and spanned 100~s. For the \gls{pcmri} results on the other hand, the final velocity fields are the average of six snapshots taken at a repetition/echo time of 6.9/3.586~ms. These differences in the averaging process stemming from limitations of each one of these approaches can explain, to a great extent, the discrepancies observed in Fig.~\ref{fig:case02_velocity_fields}. In practice, the low data sampling frequency in \gls{piv} and \gls{pcmri} measurements operates as a low-pass filter while the limited sampling interval of the simulation is equivalent to a high-pass filter. This in turn can explain the thicker and smoother shear layers observed in the former two.
The subsequent quantification presented in Fig.~\ref{fig:case02_velocity_profile} provides a better insight in occurring differences and further confirms the previously discussed effect regarding the discrepancies coming from data acquisition. Here, velocity magnitudes are shown along three orthogonal lines through the aneurysm. Overall, the courses of all approaches are in a good agreement and the values show no strong deviations from each other. It must be noted that error bars in Fig.~\ref{fig:case02_velocity_profile} only represent the 99.7\% confidence interval. The simulation results fall within the 99.7\% confidence interval at all points.\\
Matching coordinates and positions of the corresponding velocity fields, together with possible distortions in space resulting from the different experimental procedures, constitute also a major challenge for this comparison; it certainly explains to some extent the observed discrepancy. This effect is particularly visible on the line along the $y$-axis in Fig.~\ref{fig:case02_velocity_profile}. While the velocity profiles from the simulation tend towards zero near the aneurysm sac walls, \gls{piv} results do not show such a clear behavior. Interestingly, the agreement between all methods is excellent for plane III (see Fig.~\ref{fig:case02_velocity_profile}, right), while larger discrepancies are noticeable in the other two (e.g., smoother courses of the measurement compared to the simulation).
Overall, while a good qualitative agreement between the three different approaches can be observed, non-negligible discrepancies persist. In-depth analysis of the results showed that the unsteady nature of the configuration, along with the different data acquisition modes ({i.e.} frequency and time-span of velocity field averaging process) affect the final velocity fields. This effect was illustrated by comparing instantaneous and averaged velocity fields in the simulation. The distortions and manufacturing artifacts in the final aneurysm geometry can also contribute to these discrepancies. Further reasons are discussed in Section~\ref{sec:discussion}.
\subsection{Under-resolved simulation of the patient-specific configuration}
\begin{figure}[!ht]
	\centering
	\hspace{-0.1\textwidth}
	\begin{subfigure}{0.3\textwidth}
		\includegraphics{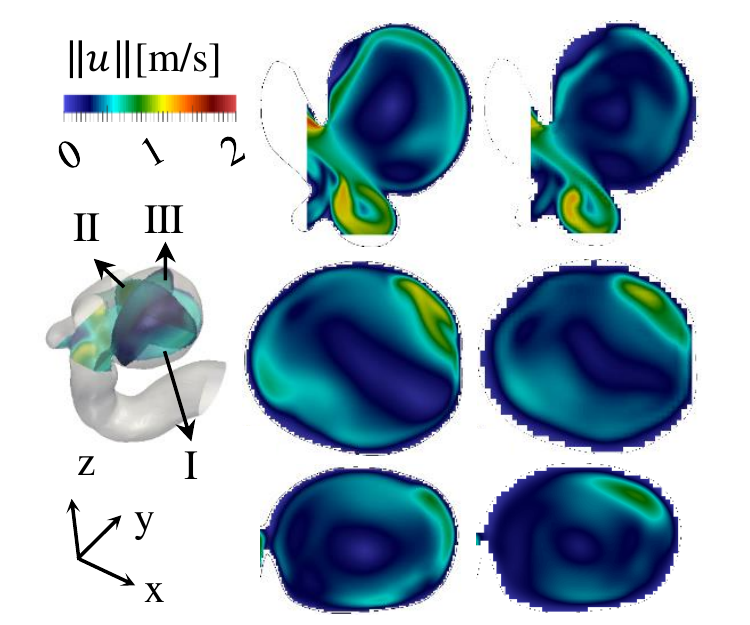}
	\end{subfigure}
	\caption{Qualitative comparison of the velocity fields in the aneurysm sac on three different planes, {i.e.} from top to bottom: planes I, II and III, as obtained from the high-resolution (left) and low-resolution (right) simulation.}
	\label{fig:low_resolution_simulation}
\end{figure}

\begin{figure}[!ht]
	\centering
	\hspace{-0.2\textwidth}
	\begin{subfigure}{0.3\textwidth}
		\includegraphics[trim={20 20 240 80},clip]{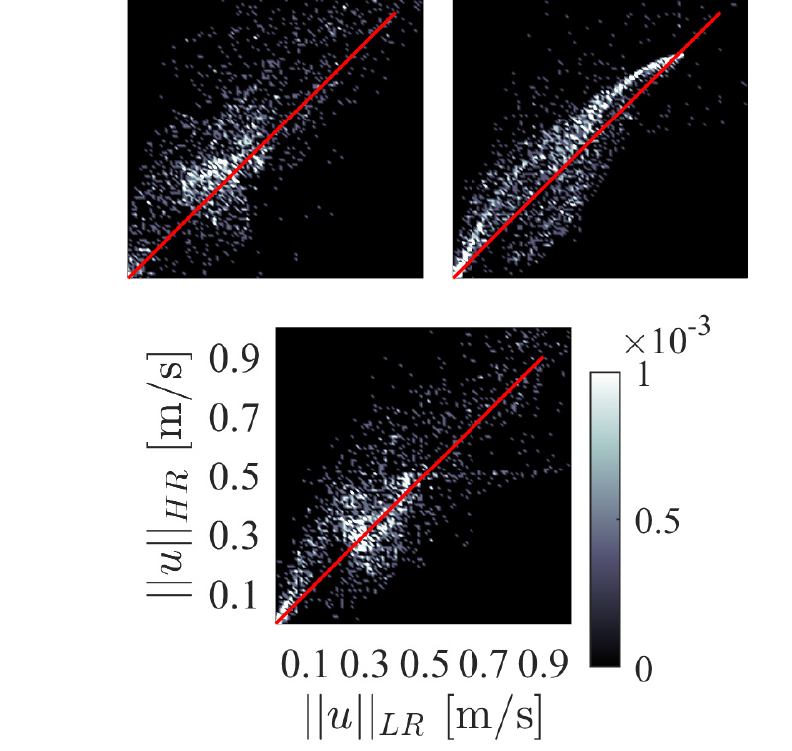}
	\end{subfigure}
	\caption{Quantitative comparison of the velocity magnitude distribution in the high- and low-resolution simulations on planes I (top left), II (top right) and III (bottom). The color bar represents the normalized (via the total number of sampled points) number of grid-points with the corresponding velocities in the low-resolution ($x$-axis) and high-resolution ($y$-axis) simulation.}
	\label{fig:VMSI}
\end{figure}

\begin{figure}[!ht]
	\centering
	\hspace{-0.2\textwidth}
	\begin{subfigure}{0.3\textwidth}
		\includegraphics[trim={10 20 180 140},clip]{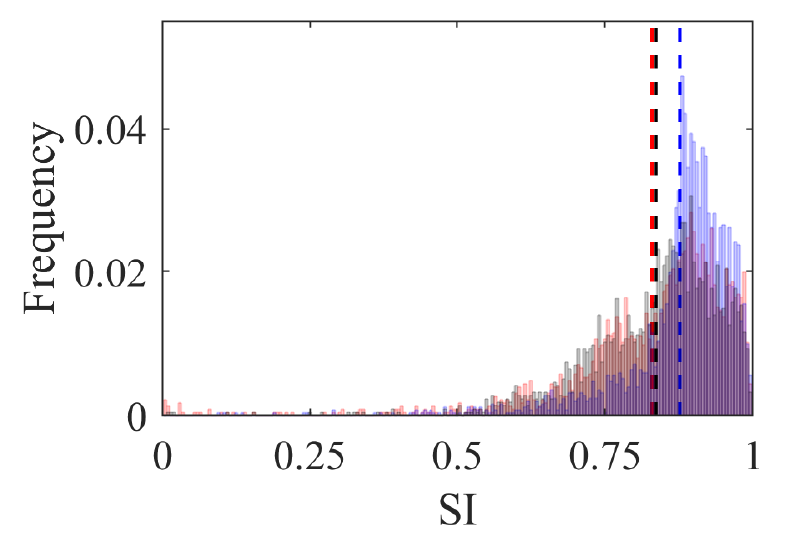}
	\end{subfigure}
	\caption{Histograms displaying the normalized frequency of the SI on planes I (blue), II (red) and III (black). The corresponding overall SIs are shown using dashed lines of the same color.}
	\label{fig:SI}
\end{figure}

Finally, to showcase the ability of the collision operator to deal with under-resolved cases, and provide a more in-depth analysis, the patient-specific configuration was also modeled using a coarse grid. The time-averaged velocity fields obtained using the high- and the low-resolution \gls{lbm} simulation are shown in Fig.~\ref{fig:low_resolution_simulation}.
The orthogonal planes reveal that the predicted flow structures, while admitting discrepancies, agree very well. The quantitative comparison of both simulations leads to a domain-averaged similarity of 0.8294, 0.8765 and 0.8359 for planes I, II and III, respectively. This similarity index on each plane is computed as the average similarity index over all discrete points on the plane \cite{gaidzik2019transient}:
\begin{multline}
    \hbox{SI} =  \left(1+\frac{\bm{u}_{\hbox{\footnotesize HR}}\cdot\bm{u}_{\hbox{\footnotesize LR}}}{\lvert\lvert \bm{u}_{\hbox{\footnotesize HR}} \lvert\lvert \lvert\lvert \bm{u}_{\hbox{\footnotesize LR}} \lvert\lvert}\right) \\ \left(1 - \lvert \frac{\lvert\lvert \bm{u}_{\hbox{\footnotesize HR}} \lvert\lvert}{\lvert\lvert \bm{u}_{\hbox{\footnotesize HR}} \lvert\lvert_{\hbox{\footnotesize max}}} - \frac{\lvert\lvert \bm{u}_{\hbox{\footnotesize LR}} \lvert\lvert}{\lvert\lvert \bm{u}_{\hbox{\footnotesize LR}} \lvert\lvert_{\hbox{\footnotesize max}}}\lvert\right),
\end{multline}
where $\bm{u}_{\hbox{\footnotesize HR}}$ and $\bm{u}_{\hbox{\footnotesize LR}}$ are the velocity vectors from the high- and low-resolution simulations, respectively. The deviations are further illustrated in Fig.~\ref{fig:VMSI}, where the low-resolution velocity magnitudes are plotted against the high-resolution ones on the three considered planes. The normalized frequency of the similarity index on all three planes is shown in Fig.~\ref{fig:SI}.
Despite these minor differences with respect to the velocity distribution, it is important to mention that the low-resolution simulation reduced the number of grid-points by a factor of eight, and -- given the acoustic scaling -- the number of time-steps needed for convergence by a factor of two. This in turn reduced the computation time by a factor of 14 on the same machine and using the same number of processing units.
\section{Discussion} \label{sec:discussion}
With an improved linkage of medical and engineering disciplines as well as increasing computational resources new potentials in interdisciplinary research are revealed. Specifically, the challenging question of rupture risk assessment for intracranial aneurysm remains unanswered until now; however, more and more insights into patient-specific hemodynamics become accessible~\cite{Murayama.2019}. On the other hand, insufficient data processing, the absence of reliable boundary conditions and mostly long simulation times lead to a reduced translation into clinical practice~\cite{Berg.2016}.\\
To highlight the potential of image-based blood flow simulations and demonstrate its reliability with respect to flow prediction, a multimodal validation study was carried out. This included different configuration, flow scenarios, measurement methodologies and simulation settings, respectively.\\
The first part of the validation relied on flow assessments under well-controlled conditions. A brief comparative study showed that the \gls{chmrt} collision operator with a full extension of the \gls{edf} performed much better than the classical \gls{srt} model. To further validate the solver an idealized model of an intracranial aneurysm was created leading to the development of an organized flow field. The qualitative and quantitative comparison of both independent techniques (\gls{lbm} versus \gls{piv}) showed an excellent agreement. Due to the unsteady inflow, minor fluctuations occurred throughout the cardiac cycle, especially near peak systole. Still, both simulation and measurement were in very good agreement during the whole period.\\
In a second part of this study, complexity was increased by considering a patient-specific anatomy. Although unsteady phenomena occurred at peak-systolic inflow conditions, flow predictions by simulation and measurement were comparable. Beside the validation using \gls{piv}, \gls{pcmri} acquisitions were carried out. Here, minor differences especially in high-velocity regions close to the aneurysmal lumen were present, which could be a result of vessel wall compliance~\cite{tupin2020wallcompliance}. However, taking into account all underlying differences with respect to temporal and spatial resolution as well as the uncertainties associated to the measurement procedures, the validation of the \gls{lbm} simulation was successfully conducted.\\
Apart from the validation purpose, additional simulations with low spatial resolutions were performed. The comparison of low- and high-resolution \gls{lbm} computation showed that similar velocity fields were calculated, although only $1/14$ of the simulation time was required. Hence, depending on the clinical research question and the desired accuracy of the hemodynamic description the low-resolution approach, made possible by the \gls{chmrt} collision operator, can be a promising development direction, especially given that the computational overhead associated to the change of collision  operator is very minor~\cite{coreixas2019comprehensive}.\\
Beside the presented findings, it is important to mention that this study has several limitations: 
First, the number of cases considered is relatively small for an extensive validation study. Only two idealized and one patient-specific geometries were selected. However, it is important to point out that intracranial aneurysms show a high variability regarding size (small versus giant), location (lateral versus terminal), morphology (spherical versus complex). Consequently, this can lead to increased challenges for the underlying numerical schemes. Nevertheless, multiple numerical and experimental processing steps are involved.\\
Second, precise qualitative and quantitative comparisons are only feasible, when identical conditions are warranted. However, due to the manufacturing of the phantom models for the in-vitro investigations or the subsequent registrations processes, minor misalignment between the ideal CAD geometries and the silicone models can occur.\\
Third, blood was considered as an incompressible Newtonian fluid in  the simulations. There is an ongoing debate about the necessity of a more detailed consideration of this suspension~\cite{saqr2019rheology,Lee.2019,Mahrous.2020}. Nevertheless, blood rheology still appears to have a secondary impact compared to primary factors such as segmentation or boundary conditions~\cite{Khan.2017,Berg.2019}.
Finally, although \gls{pcmri} is a technique to obtain flow data in-vivo, within this study in-vitro measurements in a silicone phantom were carried out. Therefore, the MR sequence might lead to slightly different results when applied to a real aneurysm. \\
{Since this study demonstrates the feasibility of applying the introduced \gls{lbm}-based solver to clinically relevant research problems, future work includes the consideration of an increased number of aneurysm patients. Furthermore, specific questions with respect to rupture risk assessment, thrombus formation or treatment support will be addressed, while accompanying experimental validation measurements are carried out to ensure the reliability of our numerical models.}
\section{Conclusions}
\newabbreviation{mlups}{MLUPS}{Mega lattice updates per second}
This validation study comprising multimodal measurement techniques and various experimental scenarios demonstrates that the presented \gls{lbm} solver relying on a \gls{chmrt} collision operator is able to generate reliable and valid simulation results. These results are in line with observations made in the \gls{lbm} literature on the effect of such models in eliminating a number of numerical artefacts associated to the classical \gls{srt} collision operator with second-order \gls{edf} {while maintaining a low computational cost low -- around 0.4 \gls{mlups} per processor. As shown in~\cite{HosseiniPhD2020} the added cost of the modified collision operator is negligible.}
\newabbreviation{les}{LES}{large eddy simulations}
Furthermore, the wider stability domain and enhanced spectral properties of the collision model (as compared to the \gls{srt}) allowed for simulations relying on lower resolutions (not stable with the \gls{srt}) yielding comparable numerical predictions with an extensive reduction of computational load (grid-points and time-steps). Given the difficulties of applying explicit sub-grid model-based \gls{les} to pulsatile flows in complex geometries, tools and numerical formulations allowing for model-free under-resolved simulations -- tantamount to implicit \gls{les} -- can be an interesting path towards a rapid evaluation of flow parameters in complex biomedical settings.
\section*{Acknowledgements}
S.A.H. would like to acknowledge the financial support of the Deutsche Forschungsgemeinschaft (DFG, German Research Foundation) in TRR 287 (Project-ID 422037413).\\
F.H. was supported by the State Scholarship Fund of the China Scholarship Council (grant number 201908080236).\\
Furthermore, P.B. acknowledges the funding by the Federal Ministry of Education and Research within the Research Campus \textit{STIMULATE} ({13GW0473A}) and the German Research Foundation (BE 6230/2-1). \\
The authors further thank Franziska Gaidzik and Daniel Stucht (both University of Magdeburg, Germany) for their assistance during the \gls{pcmri} flow measurements.
\bibliography{references}
\end{document}